# Reading hidden writing and drawings on papyrus using speckle optical technique and multispectral images.

**L. Buffarini[1-3], H. J. Rabal[1], N. L. Cap[1], E.E. Grumel[1-2], M. Trivi[1-2], M. Tebaldi[1-2]**

[1]*Centro de Investigaciones Ópticas (CONICET La Plata – CIC- FI UNLP)*
[2]*Dpto. Cs. Básicas, Fac. Ingeniería, Universidad Nacional de La Plata*
*Camino Parque Centenario e/505 and 506, M. B. Gonnet (1897), La Plata, Argentina.*
[3] *Facultad Regional Avellaneda. Universidad Tecnológica Nacional.*

**ABSTRACT**

We carried out an experimental set-up for reading a hidden drawing under papyrus using an optical technique based on the speckle phenomenon that is observed when a rough surface is illuminated by laser light. We propose the use of several wavelengths of light illumination and adaptive algorithms to process the speckle images. We employ a set of filters to produce better discrimination of the results as visually judged. This approach is a cheap and relatively simple non-destructive and non-invasive procedure that could be extended to other painted objects or subsurface hidings of archaeological interest, as well as, to other dynamic speckle experiments.

**KEYWORDS:** ARCHEOLOGICAL PAINTED OBJECTS, EGYPTIAN PAPYRUS, DYNAMIC SPECKLE, MULTISPECTRAL IMAGE PROCESSING, SUBSURFACE READING.

# 1. INTRODUCTION

The reading of hidden writings or the observation of drawings under paper or papyrus is a subject that is of interest in Archaeology because it allows one to access hidden information without damaging the document [1,2]. The reading of writing or drawing ancient documents is very important in History [3,4]. Egyptians (and other early cultures) carefully rolled the papyri to occupy the lesser available space. To read them they had to be very carefully unfolded. Nowadays, these papyri have considerably aged so that valorous texts could be easily damaged if we intend to unfold or unroll them [1].

Many efforts have been developed for this purpose, but destructive methods cannot be used. So, non-destructive methods should then be required, including synchrotron radiation, optical techniques (multispectral imaging with reflection and transillumination, and optical coherence tomography), X-ray (X-ray fluorescence imaging, X-ray fluorescence spectroscopy, X-ray micro-computed tomography and phase contrast X-ray) and terahertz-based approaches. [See Ref 2 and references therein]. However, these techniques use expensive instruments and in some cases are difficult to perform. Therefore, they are not easily available in the Archaeology and Heritage environments and interdisciplinary work is necessary to access them.

Thus, developing cheap and non-intrusive techniques is an interesting field to explore. An approach using low-cost and easily accessible optical methods would be useful for extended use in museums and archives.

This operation could be easy to test non-destructively in repeatable situations with comparatively inexpensive instruments if we know in advance in the tests what the result should resemble. Then the investigation could be extended to samples whose content was unknown and its validity checked with alternative techniques.

Optical speckle techniques bear in mind these requirements and may be an option that can be explored [5,6]. On the other hand, multispectral image processing has received considerable attention due to its important applications in medicine, biology, remote sensing, communications, TV displays, etc. [7,9].

## 2. RESEARCH AIM

To preserve the objects from cultural heritage for future generations, there is a need to study samples non-destructively. In this sense, we present an approach that combines the speckle technique with multispectral images to study objects of historical relevance. We propose a speckle image method obtained with several wavelengths of the illumination laser beam for reading under hiding media such as paper and papyrus. The proposed speckle technique is a non-invasive, low-cost, and easily accessible alternative to studying objects of cultural heritage. To test the validity of the proposal, we evaluate a mock papyrus replica with the results well known beforehand.

## 3. METHODS

Dynamic speckle is an optical technique widely used in multiple applications [10,11]. Many algorithms have been proposed to describe it [12,13] that provide high-quality results for some applications (seeds, fruits, bacteria, etc.). There is not a single algorithm that is useful for all applications. Usually several of them must be tested for a certain measurement. Recently, we proposed [14] a set of new tuneable algorithms for screening activity speckle images, an improvement on current algorithms, and a new one, all of them based on a generalization of the addition and subtraction complex operations.

In our experiment here, we applied tuneable algorithms to multi-wavelength laser speckle images. The object is a reversed papyrus with coloured drawings that is slightly heated with an incandescent lamp to induce speckle activity.

In section 4, we are going briefly describe the speckle phenomenon and the employed algorithms to process the data. Section 5 is devoted to the experimental procedure. Section 6 deals with the results. Finally, Section 7 summarizes the conclusions.

4. **SPECKLE PHENOMENON AND TUNEABLE ALGORITHMS**

When a coherent beam coming from a laser illuminates a rough object, we observe a random interference pattern called *speckle*. [15]. Fig. 1 shows a typical speckle pattern.

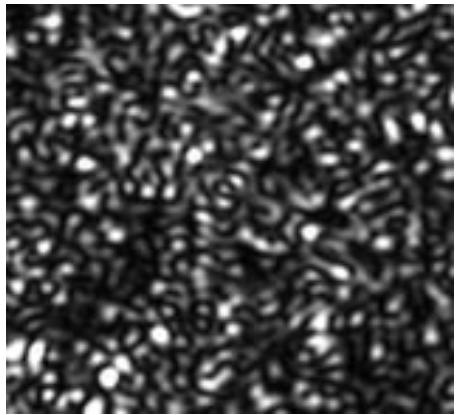

Figure 1: Typical speckle pattern

If the surface of the objects presents some type of local movement, then the intensity and the shape of the observed speckle evolve in time. The speckle patterns thus become time-dependent and contain information about the object's motion. Therefore, the study of the temporary evolution of the speckle patterns may provide an interesting tool to characterize the parameters involved in the dynamic processes of an object.

Data of many speckle images must usually be processed in an experiment to characterize an object using the dynamic speckle technique. Typically, about 4000 frames are recorded of at least 300 x 300 pixels each (0 to 255 gray levels). Different descriptors can be used to analyze them. In our case, we employ a recently reported set of new tuneable algorithms to process the speckle images [14] based on a generalization of the addition operation. In that paper, a detailed description of the algorithms used in our papyrus experiment is given (Fujii [16], Absolute Value of Differences (AVD) [17], Generalized Differences (GD) [18], and Tuneable t algorithms [14]). Also, an Appendix with a brief description of the used algorithms is included at the end of this paper.

5.     **EXPERIMENTAL SET-UP**

Speckle experiments were carried out for testing the performance of our proposal using a colour Egyptian drawing replica as observed from the unpainted surface of the papyrus. The object was located over an opaque background so that it could not be seen by translucence.

Speckle patterns are obtained when the sample under analysis is illuminated with laser light depending on the wavelength of the light used. Then, in our proposal, the sample was illuminated with different wavelengths. We used four different expanded laser sources to illuminate the papyrus: a blue semiconductor diode laser with a wavelength of 473 nm and a maximum nominal power of 50 mW, a green semiconductor diode with 532 nm wavelength of and a nominal power maximum of 150 mW, a low-power He-Ne red laser with a wavelength of 632.8 nm and a nominal maximum power of 30 mW and an infrared semiconductor diode laser, 780 nm wavelength and a nominal maximum power 30 mW. These illuminations did not produce any damage to the object. The sample was slightly heated by an incandescent lamp to induce speckle activity. The temperature rise due to the heating lamp was 3ºC which is much less than the daily meteorological excursion. During register, the lamp was turned off.

A CCD camera (Pulnix interline transfer, TM-6CN model, with 8.6 $\mu$m x 8.3 $\mu$m cell size) was used to register a series of 4000 speckle images for each illumination wavelength. They were digitized by a frame grabber, stored, and processed by a personal computer. Fig. 2 shows the experimental setup. The use of multiple wavelengths offers comparative advantages in comparison with a single laser. They give several results due to the different penetration of light into matter, the different reflectivity of colour drawings, and the dependence of the speckle activity measured on the wavelength.

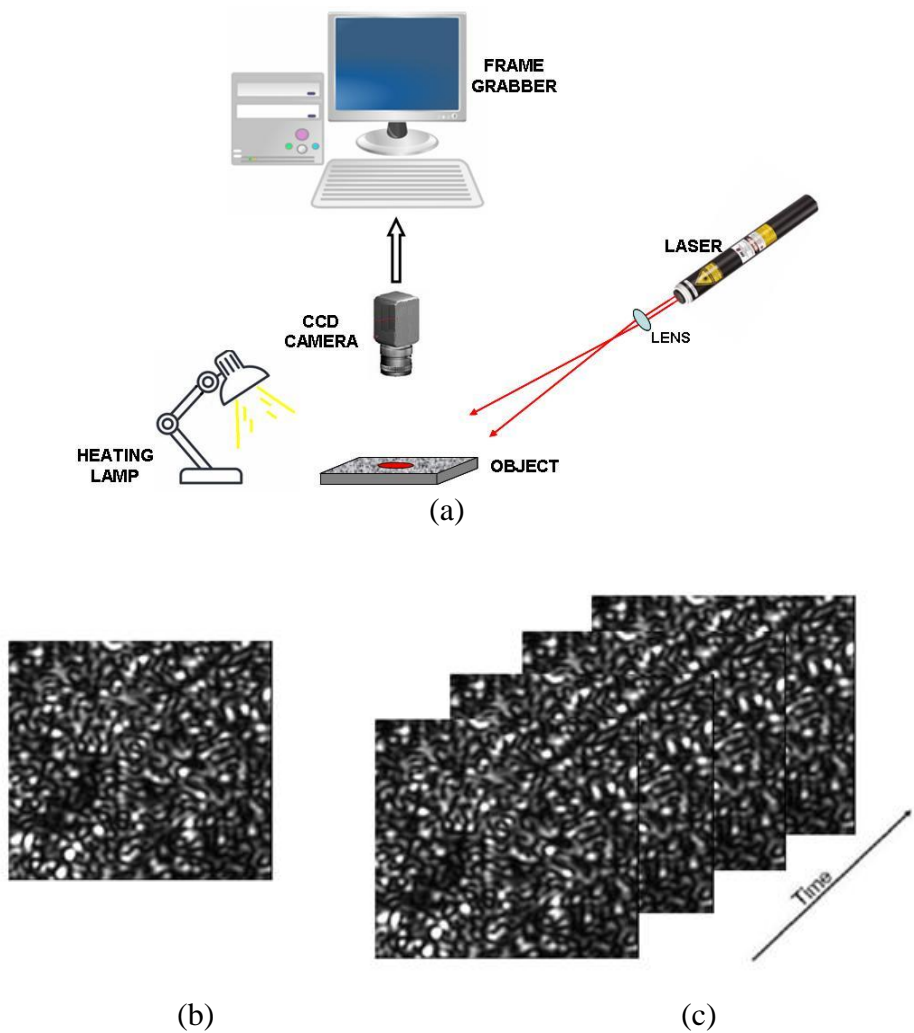

Figure 2: (a) Experimental set-up. (b) speckle pattern of a single frame. (c) stack of successive speckle frames

## 6. EXPERIMENTAL RESULTS: PAPYRUS SPECKLE IMAGES

The object consisted of a replica of an Egyptian drawing painted on the original papyrus as shown in Fig. 3.

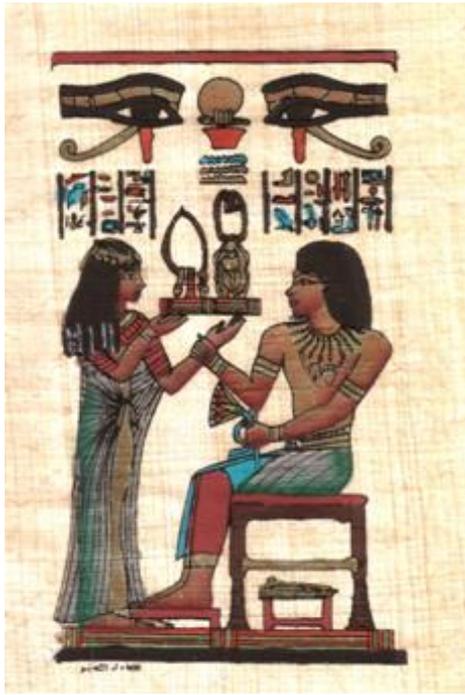
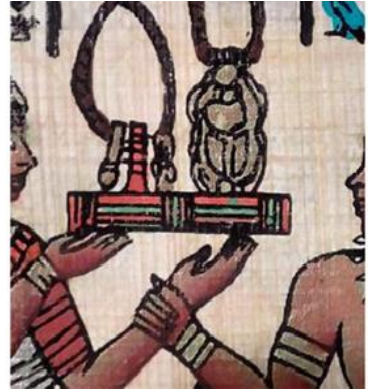

(a) (b)

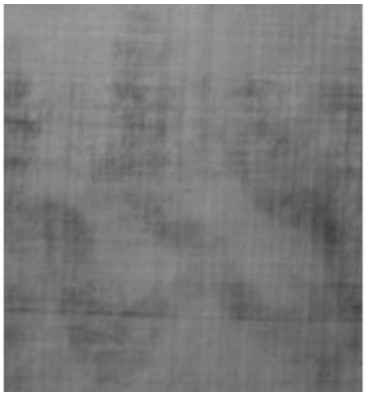
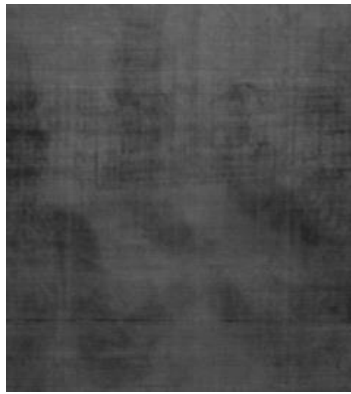
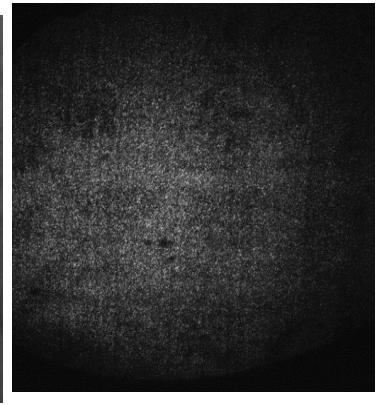

(c) (d) (e)

Figure 3: (a) Image of a replica painted over papyrus (Original copy), (b) the chosen region, (c) image under incoherent light illumination, copy turned over, (d) as seen in near infrared light and (e) under laser illumination (unprocessed speckle image).

Figure 3 (a) shows an image of the sample: a replica painted over papyrus. Fig. 3 (b) shows the chosen region used for processing. Fig. 3 (c) shows the copy turned over as it is seen under ordinary incoherent illumination. Fig. 3 (d) how it is seen in near-infrared light and Fig. 3 (e) speckled image as observed when illuminated by laser visible light.

Notice that in Fig. 3 (c), colour is lost, and details are blurred. Also, in Fig. 3 (d) it is still blurred and in Fig. 3 (e) the speckle image obtained with laser illumination shows no hint of the hidden drawing. This speckle image does not permit recognition of the drawing.

It should be mentioned that for speckle images obtained with each laser and parameters a huge number of images can be constructed with the algorithms. Here we show the visually judged best results so far.

The results of processing the speckle images using different algorithms and the proposed lasers are shown below.

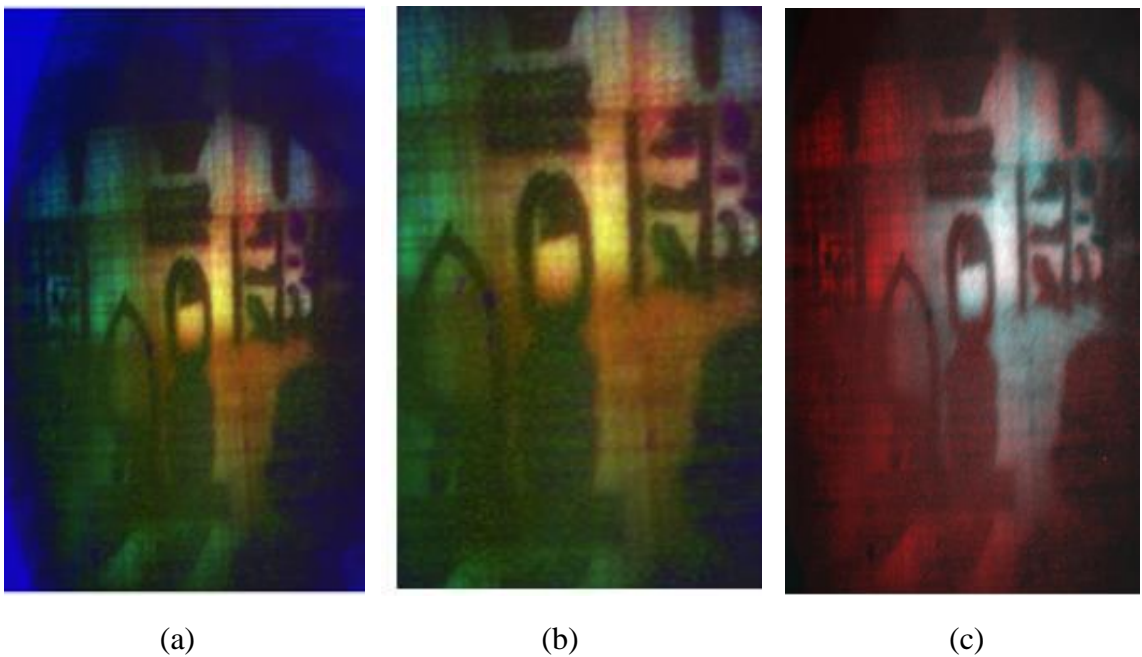

(a) (b) (c)

Figure 4: (a) Component R with AVD ($\varphi = 5$) in red laser light. Component G with AVD ($\varphi = 5$) in green laser and component B corresponds to Tuneable $\tau$ ($\varphi = 80$) blue laser; (b) cut-out and enlargement of Fig. 4 a) and (c) pseudo coloured version with AVD ($\varphi = 0$), processed with red laser and exponent LUT $\alpha=1$.

Fig. 4 shows the images of RGB (red, green, blue) composition obtained with the AVD algorithm [17] in particular; each colour channel corresponds to the results obtained using the speckle pattern with red, green, and blue lasers and with AVG algorithm: in Fig. 4 a), the results of the algorithms as formed by: component R is the results obtained with AVD

($\varphi$ = 5) in red laser light; component G is the result obtained with AVD ($\varphi$ = 5) in green laser light and component B corresponds to τ ($\varphi$ = 80) with blue laser illumination. Fig. 4 (b) is a cut-out and enlargement version of Fig. 4 (a). Fig. 4 (c) shows a pseudocoloured version of the stack processed with AVD ($\varphi$ = 0) with red laser and exponent LUT α=1.

It can be observed that the original object is recognizable in Fig. 4 (a). Details are better observed in Fig. 4 (b) and 4 (c). The chosen angle in Fig 4 (c) is $\varphi$ = 0 because as the angle increases the result worsens. In addition, the blue one detects edges with a smaller blur than the red (see Fig. 4 (c)) but the red covers a wider area. Note that the structure of the papyrus support fibers can be clearly seen in Figure 4 (a) in comparison with 4 (c).

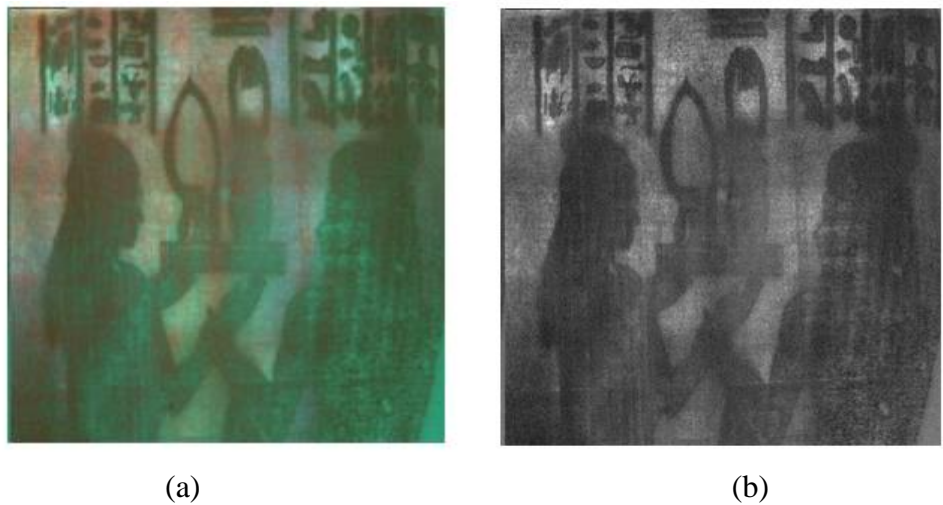

(a)                     (b)

Figure 5: (a) Processed image obtained with infrared laser using the three proposed algorithms: the red component is AVD ($\varphi$ = 0), the green component is Fujii ($\varphi$ = 110) and the B blue component is Tuneable τ ($\varphi$ = 70). (b) Processed image obtained only with Tuneable τ ($\varphi$ = 70).

Fig. 5 (a) shows the processed image obtained with an infrared laser. In this case, each colour channel was obtained using the speckle pattern corresponding to infrared laser in all cases and with different algorithms for each colour channel. The processing used the Tunable (τ), Fujii, and AVD proposed algorithms, then the best angles were chosen for each one, and an RGB was composed with the results. The red component of the final RGB result corresponds to AVD ($\varphi$ = 0), the green component corresponds to Fujii ($\varphi$ =

110) and the blue component is Tunable τ (φ = 70). Fig. 5 (b) shows the processed image obtained with an infrared speckle pattern and only with Tunable τ (φ = 70).

Notice that most of the original images can be recognized and details are better identified. Parts of the papyrus material can be also recognized and the structure of the fibers of the support can also be seen. The background intensity variations are presumably due to uneven heating.

These results compete favourably with those obtained before using Fractal Dimension [19] on the same stack of speckle images.

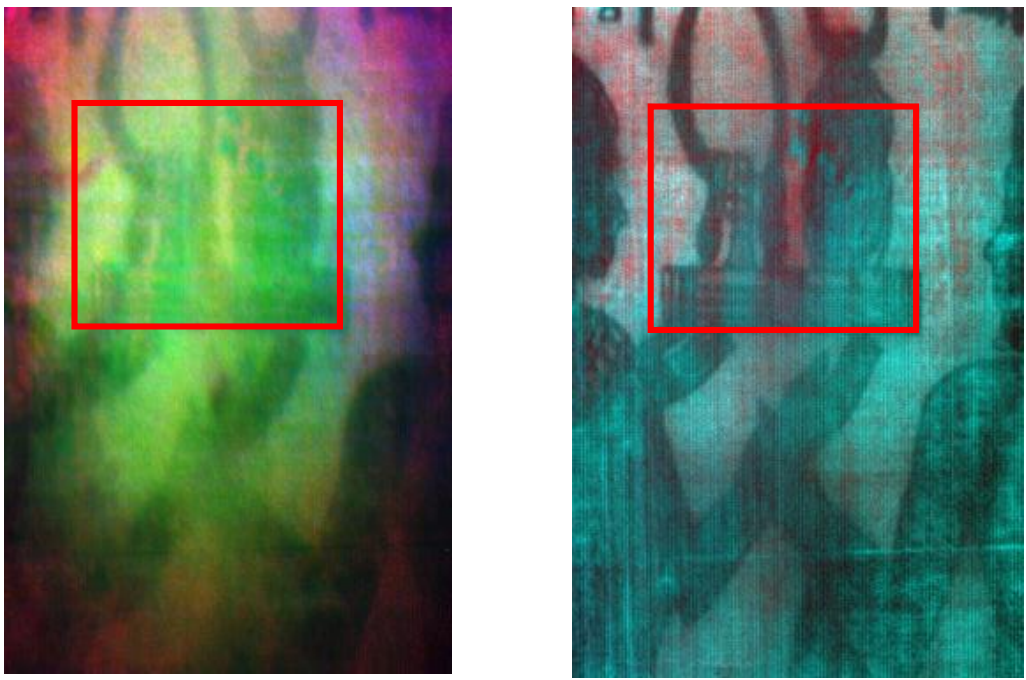

(a) (b)

Figure 6: (a) RGB image result corresponds to R component obtained with AVD (φ = 0) α=1 and infrared light, G component obtained with AVD (φ = 110) α=1 and red light, and B component obtained with AVD (φ = 70) α=1 for the blue laser illumination. (b) RGB image correspond to the best results for the three algorithms using infrared illumination. It is composed by AVD (φ =5) α=2 in the R component, Fujii (φ =50) α=2 in the G component and τ (φ =75) α=2 in the B component.

Fig. 6 shows the results of composing RGB images where each component is obtained with three laser lights (infrared, red, and blue) and three algorithms (τ, AVD and Fujii). These algorithms are briefly described in appendix section. Fig. 6 (a) show results of AVD algorithm (R components correspond to infrared speckle patten and AVD (φ = 0) α=1, G components correspond to red speckle patten and AVD (φ = 110) α=1, and B components is AVD (φ = 70) α=1 for the blue laser illumination).

In Fig. 6 (b) it is shown the composed RGB image where each component is obtained with the best results obtained from the three algorithms and for the (visually judged) best angles by using infrared illumination. It is composed by AVD (φ =5) α=2 in the R component, Fujii (φ =50) α=2 in the G component and τ (φ =75) α=2 in the B component.

The improvement of Fig. 6 (b) with respect to Figure 6 a) is clearly observed. Fig. 6 a) shows that the image contains considerable blurriness, while the image in Fig. 6 b) contains a better detailed definition. The lines within the red rectangle and the profile of the human figures are well defined. Details of the structure of the support fibers can also be seen.

Note that the proposed algorithms allow obtaining several results. Then, it is possible to choose the result showing a better description of the hidden object [14].

To verify these observations, an edge detector filter is applied. The Canny edge detection algorithm [20] uses a filter based on the first derivative of a Gaussian. Since it is susceptible to noise present in the raw image data, the original image is transformed with a Gaussian filter. In Fig. 7 (a) and 7 (b) the same edge detection filter is applied using the Canny filter with a threshold of 0.7.

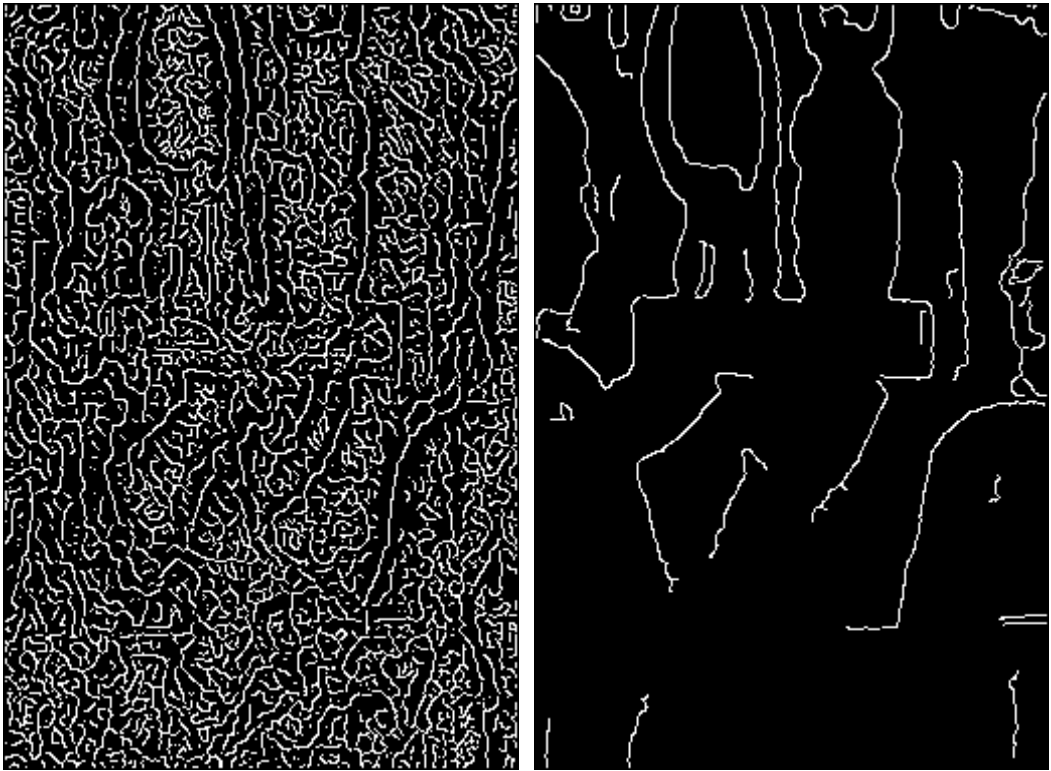

a) b)

Figure 7: a) and b) image of Figure 6 a) and 6 b) after edge detection processing

(Canny), respectively [Ref. 21].

In Fig. 7 (a), it is shown how the algorithm detects edges that correspond to the homogeneous background of the image. That is, it detects edges in excess.

In Fig. 7 (b), it detects fewer edges, and the outline of the human figures and the object in the centre are marked. In addition, in most of the areas where there is a discontinuity in the intensity of the pixels, they are detected.

# 7. CONCLUSIONS

We proposed the use dynamic speckle technique to observe hidden drawings under papyrus.

We employed lasers with several wavelengths combined with tunable speckle algorithms. The information obtained from the algorithms provides many results. This is a consequence that there is not a single descriptor that is useful for all the cases. For best visually judged results all algorithms should be tested. The procedure could be repeated for several wavelengths providing different results. These results consist of a very important number of images constituting a database. Then for the construction of an RGB image, three components must be chosen from the fore mentioned database.

Fig. 4 and 5 show some of the different combinations for the judged best results. On the other way, there are methods of quantitative evaluation of quality such as PSNR (Peak Signal-to-Noise Ratio) or edge visibility, but they are not always visually better.

Besides, the use of several colour images digitally processed is of use in DStretch processing to identify objects in painted walls in Egyptian tombs from Beni Hassan [21]. Both techniques could be combined.

Fig. 6 shows a comparison between the images obtained with different lasers (Fig. 6 (a)) and those obtained with infrared lasers (Fig. 6 (b)). In the indicated rectangle, the details are clearly recognizable in Fig. 6 (b) while they are much more diffuse in Fig. 6 (a)

The proposed method is relatively cheap, non-invasive, non-destructive, and easy to implement in comparison with other expensive techniques already well-established for these studies. It is not limited to the papyrus example presented here.

This approach has also been successfully applied to reading under sticked paper, Liquid paper®, in seeing through turbid media in incoherent illumination (inside a reactor), and through rain and fog. The results will be published elsewhere.

This procedure could also be tested, by using speckle illumination on the reverse of paintings, to look for *pentimenti*, that is, traces of earlier working that the artist himself later altered. It could be tested also on rock paintings or painted skulls or other painted or written objects of archaeological or artistic interest or in any situation where subsurface inhomogeneity detection was of interest.

**ACKNOWLEDGEMENTS**

This work was supported by Facultad de Ingeniería, Universidad de La Plata Grant I239, Agencia Nacional Promoción Científica y Tecnológica ANPCyT Grant PICT Nº. 4558, 2087and PICT-2021-CAT-I-00074. Consejo Nacional de Investigaciones Científicas y Técnicas and Comisión de Investigaciones Científicas de la Provincia de Buenos Aires, Argentina.

# APPENDIX

In this section, we briefly describe the algorithms used in this work. A more detailed description can be found in Reference [14].

Let $I^k(i,j)$ indicate the value of the recorded intensity at pixel $(i,j)$ in the *k-th* frame of the dynamic speckle stack corresponding to N images.

Then, we define the $AVD(\varphi)$ [1] as:

$$AVD_{i,j}(\varphi) = \frac{1}{N-1}\sum_{k=1}^{N-1} \sqrt{\left|(I_{i,j}^k)^2 + (I_{i,j}^{k+1})^2 + 2I_{i,j}^k I_{i,j}^{k+1} \cos(\pi-\varphi)\right|} \qquad (1)$$

The $Fujii(\varphi)$ measure [2] is defined as:

$$Fujii_{i,j}(\varphi) = \sum_{a,b=1}^{N-1} \sqrt{\left|\frac{(I_{i,j}^k)^2+(I_{i,j}^{k+1})^2+2I_{i,j}^k I_{i,j}^{k+1}\cos(\pi-\varphi)}{(I_{i,j}^k)^2+(I_{i,j}^{k+1})^2+2I_{i,j}^k I_{i,j}^{k+1}\cos(\varphi)}\right|} \qquad (2)$$

This algorithm gives high quality results when the image is uniformly illuminated [11].

Finally, we define the $\tau(\varphi)$ algorithm [3] as:

$$\tau(\varphi)_{i,j} = \sum_{a,b,c=1}^{N-2} \sqrt{\left|\left(\frac{\frac{(a_{i,j}^2+b_{i,j}^2+2a_{i,j}b_{i,j}\cos(\pi-\varphi))(a_{i,j}^2+c_{i,j}^2+2a_{i,j}c_{i,j}\cos(\pi-\varphi))}{(b_{i,j}^2+c_{i,j}^2+2b_{i,j}c_{i,j}\cos(\pi-\varphi))}}{\frac{(a_{i,j}^2+b_{i,j}^2+2a_{i,j}b_{i,j}\cos(\varphi))(a_{i,j}^2+c_{i,j}^2+2a_{i,j}c_{i,j}\cos(\varphi))}{(b_{i,j}^2+c_{i,j}^2+2b_{i,j}c_{i,j}\cos(\varphi))}}\right)\right|} \qquad (3)$$

We illustrate the expression of the proposed tuneable algorithm for a simple case of operating on 3 consecutive frames and averaging the results. We call $a, b, c$ the gray level values taken by a pixel $(i,j)$ in every set of three consecutive frames.

All these modified algorithms have a free parameter, namely angle $\varphi$ that can be chosen by the user to obtain the better visual results.

As the results of the aforementioned algorithms exhibit a high number of possible numerical outputs that cannot be displayed in usual monitors, we also use a radiometric transformation to choose the enhancement of different regions of the histogram.

$$I_{output} = I_{input}^{\alpha}$$

where $I_{input}$ is the calculated value with the mentioned algorithms and $I_{output}$ is the value effectively shown in the monitor and α is chosen by the user to enhance or reduce different parts of the histogram.